\documentclass[11pt]{article}

\usepackage{amssymb,bm,amsmath}
\usepackage{color, graphicx}
\usepackage{multirow}
\usepackage{booktabs}
\usepackage{mathtools}

\newcommand{\Z}{\bf{Z}}

\begin{document}

\begin{titlepage}

\begin{center}
{\Large\bf Charting the scaling region of the Ising universality class in two and three dimensions}
\end{center}
\vskip1.3cm
\centerline{Michele Caselle$^1$ and Marianna Sorba$^{1,2}$}
\vskip1.5cm
\centerline{\sl $^1$ Department of Physics, University of Turin \& INFN, Turin}
\centerline{\sl Via Pietro Giuria 1, I-10125 Turin, Italy}
\vskip0.5cm
\centerline{\sl $^2$ SISSA \& INFN Sezione di Trieste}
\centerline{\sl Via Bonomea 265, 34136 Trieste, Italy}

\begin{center}
% {\sl  E-mail:} \hskip 1mm \href{mailto:caselle@to.infn.it}{{\tt caselle@to.infn.it}}, \href{mailto:anada@to.infn.it}{{\tt anada@to.infn.it}}
\end{center}
\vskip1.0cm
\begin{abstract}
We study the behaviour of a universal combination of susceptibility and correlation length in the Ising model in two and three dimensions, in presence of both magnetic and thermal perturbations, in the neighbourhood of the critical point.
In three dimensions we address the problem using a parametric representation of the equation of state. In two dimensions we make use of the exact integrability of the model along the thermal and the magnetic axes.
Our results can be used as a sort of ``reference frame" to chart the critical region of the model. 

While our results can be applied in principle to any possible realization of the Ising universality class, we address in particular, as specific examples, three  instances of Ising behaviour in finite temperature QCD related in various ways to the deconfinement transition.  
In particular, in the last of these examples, we study the critical ending point in the finite density, finite temperature phase diagram of QCD. In this finite density framework, due to well know sign problem, Montecarlo simulations are not possible and thus a direct comparison of experimental results with QFT/Statmech predictions like the one we discuss in this paper may be important.  Moreover in this example it is particularly difficult to disentangle  ``magnetic-like" from ``thermal-like" observables and thus an explicit charting of the neighbourhood of the critical point can be particularly useful. 
\end{abstract}

\end{titlepage}

\section{Introduction}

Despite its apparent simplicity the Ising model is one of the cornerstones of modern Statistical Mechanics. Over the years it has become a theoretical laboratory to test new ideas, ranging from Symmetry Breaking to Conformal Field Theories. Moreover, thanks to its exact solvability in two dimensions \cite{onsager1944,mccoy1973two} and the ease with which it can be simulated in three dimensions, it has been widely used as a benchmark to test new numerical approaches and innovative approximations. 
 
The corresponding universality class, in the Renormalization Group sense \cite{pelissetto-vicari}, is of central importance in theoretical physics due to its many experimental realizations in different physical contexts, ranging from condensed matter to high energy physics. 
At the same time  it describes the 
critical behaviour of a lot of different spin models and, in the limit of high temperatures, also of gauge theories.
 
From a Statistical Mechanics point of view it represents the simplest way to describe 
systems with short range interactions and a scalar order parameter (density or uniaxial magnetization) which undergo a symmetry breaking phase transition.
From a Quantum Field Theory (QFT) point of view it is the simplest example of a unitary Conformal Field Theory (CFT) \cite{Belavin:1984vu} perturbed by only two  relevant operators: the ``spin" operator (which is $\Z_2$ odd) and the ``energy" operator ($\Z_2$ even)~\cite{fonseca2003ising}.

Thanks to integrability, conformal perturbation and bootstrap \cite{Kos:2016ysd, berg1979} lots of results are known, both in two and in three dimensions, on the behaviour of the model at the critical point, or when only one of the two perturbing operators is present.
However, typically, the interesting regime for most of the experimental realizations of the model is when both the perturbing operators are present and much less is known in this situation.  

The aim of this paper is to partially fill this gap by studying a suitable universal combination of thermodynamic quantities (see below for the precise definition)
in presence of both perturbing operators. In three dimensions we shall address the problem using a parametric representation of the equation of state~\cite{campostrini1999}, while in two dimensions we shall make use of the exact integrability of the model in presence of a single perturbation~\cite{fonseca2003ising}.
Using these tools we shall be able to predict the value of this quantity in the whole phase space of the model in the neighbourhood of the critical point.
These values can be used as a sort of ``reference frame" to chart the critical region of the model. \\
The universal combination that we shall study involves the magnetic susceptibility and thus our proposal is particularly effective when the model is characterized by an explicit $\Z_2$ symmetry. When this is not the case, like for the liquid-vapour transition or for the finite density QCD example that we shall discuss below, the explicit knowledge of our universal combination may help to identify the exact directions in the phase space of the model with respect to which the magnetic susceptibility must be evaluated.

Thanks to universality, our results hold not only for the standard nearest neighbour Ising model, but also 
for any possible realization of the Ising universality class and in fact we shall use the high precision Montecarlo estimates obtained from an improved version of the Ising model to benchmark and test our results \cite{Hasenbusch:1998gh,Hasenbusch:1999mw,Hasenbusch:2010ua,Hasenbusch:2011yya,Hasenbusch:2015iat,Hasenbusch:2017yxr,engels2003}.

In particular, we shall concentrate in the second part of the paper on realizations in the context of high energy physics, suggested by the lattice regularization of QCD. We shall discuss three instances of Ising behaviour in finite temperature QCD related in various ways to the deconfinement transition.  In the last of these examples, we shall address the critical ending point of finite density QCD.  In this case, due to well know sign problem, Montecarlo simulations are not possible and thus a direct comparison of experimental results with QFT/Statmech predictions like the one we discuss in this paper may be important.\\

This paper is organized as follows. Sect. 2 is devoted to a general introduction to the model and to the universal combination of thermodynamic quantities which is the main subject of the paper.  In sect. 3 we shall address the problem in three dimensions using a suitable parametric representation of the equation of state of the model.  We shall also show that the same approach cannot be used in two dimensions. In sect. 4 we shall then address the two dimensional case using appropriate expansions around the exact solutions of the model. 
Finally sect. 5 will be devoted to the discussion of a set of examples in high temperature QCD. We collected in the appendices some additional material which may be useful to reproduce our numerical analysis.

\section{General information on the Ising universality class}

The Ising model has a global $\Z_2$ symmetry and is characterized by two relevant operators which encode the $\Z_2$ odd ($\sigma$) and $\Z_2$ even ($\epsilon$) perturbations of the critical point. 

From a QFT point of view, the model in the vicinity of the critical point can be written as a perturbed Conformal Field Theory 

\begin{equation}
     S = S_{CFT}+t\int d^d x \,\epsilon(x) +H\int d^d x \,\sigma(x)
     \label{QFT}
\end{equation}
where $\epsilon(x)$ and $\sigma(x)$ are the energy and spin operators and  represent 
the continuum limits of the lattice operators $\sum_{\langle ij\rangle} \sigma_i \sigma_j$ and $\sum_i\sigma_i$ respectively. 
These operators are conjugated to the reduced temperature $t=\frac{1}{T_c}(T-T_c)$ and  magnetic field $H$, which measure the deviation from the critical point. The action $S_{CFT}$ is the conformal-invariant action of the model at the critical point. In two dimensions this is the action of a free massless Majorana fermion with central charge $c=1/2$.

Thanks to the exact integrability of the model for $H=0$ (pure thermal perturbation) and for $t=0$ (pure magnetic perturbation) much is known of this QFT in two dimensions. In particular all the critical exponents and the universal ratios are know exactly and, as we shall see below, reliable expansions around the integrable lines can be constructed for several observables.\\
In three dimensions there are not exact results, but from the recent progress of the bootstrap approach and the improvement of Montecarlo methods several universal quantities can be evaluated with very high precision.

The most important realization of this QFT is the spin Ising model on a cubic (in $d=3$) or square (in $d=2$) lattice, which we shall use in the following to fix notations. As it is well known, the model is defined by the following energy function
\begin{equation}
    E(\{\sigma_i\})=-J\sum_{\langle ij\rangle} \sigma_i \sigma_j -\hat{H}\sum_{i=1}^{N} \sigma_i
    \label{def1}
\end{equation}
where the spins $\sigma_i$ can take the values $\sigma_i=\pm 1$, the index $i$ labels the sites of the lattice, the symbol $\langle ij \rangle$ means that the sum is performed over pairs of nearest neighbour sites, $J$ is the coupling strength between spins (we assume a positive
isotropic interaction so that for all pairs of nearest neighbour spins $J_{ij}=J>0$), and $\hat{H}$ is the external magnetic field. 

The partition function of the model is 
\begin{equation}
    Z=\sum_{\{\sigma_i\}}e^{-\frac{1}{k_B T}E(\{\sigma_i\})}
    \label{def2}
\end{equation}
let us define $\beta=J/k_BT$ and $H=\hat{H}/k_BT$. For $H=0$ the model is explicitly $\Z_2$ symmetric and is  characterized by two phases,  a low temperature phase in which the $\Z_2$ symmetry is spontaneously broken and a spontaneous magnetization is present and a high temperature phase in which the $\Z_2$ symmetry is restored. The two phases are separated by a critical point. If one switches on the magnetic field it becomes apparent that the low $T$ phase is actually a line of first order phase transitions which ends with the critical point. In the following we shall be interested in the scaling region in the vicinity of this critical point. Standard Renormalization Group arguments tell us that in this limit the irrelevant operators of the model can be neglected and the behaviour is completely described only by the two relevant operators  $\epsilon, \sigma$ and one can perform a continuum limit of the model which leads exactly to the QFT described by eq. \ref{QFT}.

From the partition function defined above it is easy to obtain all the thermodynamic observables.  In particular, following the standard notation we have for the magnetization and the magnetic susceptibility
\begin{equation}
    M=-\frac{\partial \log(Z)}{\partial H},\hskip 1cm
     \chi=-\frac{\partial^2 \log(Z)}{\partial H^2}=\frac{\partial M}{\partial H}
  \label{def3}
\end{equation}
The exponential correlation length can be extracted from the large distance decay of the spin-spin connected correlator as
\begin{equation}
    \langle\sigma(x)\sigma(0)\rangle_c \sim e^{-|x|/\xi} \quad |x|\to+\infty
    \label{def4}
\end{equation}
where $\langle\sigma(x)\sigma(0)\rangle_c\equiv \langle\sigma(x)\sigma(0)\rangle-\langle\sigma(0)\rangle^2$.\\
In several practical applications it is also useful the so called  ``second-moment" correlation length which is defined through the second moment of the spin-spin correlation function as
\begin{equation}
    \xi_{2}\equiv\left[\frac{1}{2d}\frac{\int d^dx\, |x|^2 \langle\sigma(x)\sigma(0)\rangle_c}{\int d^dx \,\langle\sigma(x)\sigma(0)\rangle_c}\right]^{1/2}
    \label{def5}
\end{equation}
and it is simpler to evaluate than $\xi$ both in numerical simulations and in experiments.

In the scaling limit the critical behaviour of all thermodynamic quantities is controlled by the two ``scaling exponents" $x_\epsilon$ and $x_\sigma$ which are universal and are shared by all physical realizations of the Ising universality class.
The corresponding amplitudes are not universal, but one can construct suitable combinations in which the non-universal features of the model cancel out (see appendix \ref{App:A}) and represent testable predictions of the Ising QFT to be compared with any possible realization of the Ising universality class.

While this is a well studied subject when only a single perturbation is present, its extension to the whole scaling region of the model, where both the $H$ and $t$  perturbations are present is not straightforward. 

The main goal of this paper is to show that such an extension can be easily obtained making use of a {\sl parametric representation} of the model and that the resulting universal quantities can be used as a natural ``reference frame" to chart the scaling region of the Ising universality class.  

While the parametric approach is completely general and could be applied in principle to any universal combination of thermodynamic quantities, in this paper we shall study in particular the following ratio
\begin{equation}
    \Omega=\left(\frac{\chi(t,H)}{\Gamma_-}\right) \left(\frac{\xi_-}{\xi(t,H)}\right)^{\gamma / \nu}
    \label{Omega}
\end{equation}
and its natural extension to the second moment correlation length

\begin{equation}
    \Omega_2=\left(\frac{\chi(t,H)}{\Gamma_-}\right) \left(\frac{\xi_{2,-}}{\xi_2(t,H)}\right)^{\gamma / \nu}
    \label{Omega2}
\end{equation}

where  $\Gamma_-$, $\xi_-$ and  $\xi_{2,-}$ denote the amplitudes of $\chi$, $\xi$ and $\xi_2$  along the $t<0,H=0$ axis (see appendix \ref{App:A} for detailed definitions and normalizations).

This choice is motivated by the fact that the two observables which appear in the ratio are rather easy to evaluate, both in numerical simulations and in experiments, since they only involve derivatives or correlations of the order parameter and are normalized with respect to the values they have along the critical line of first order phase transitions, which is easy to identify (again, both numerically and experimentally).  

The main drawback of this choice is that it assumes an explicit realization of the $\Z_2$ symmetry. While in many interesting applications, like for the liquid-vapour transition in which the role of the perturbing parameter is played by the density, this is not the case 
and the $\Z_2$ symmetry is just an ``emergent" symmetry. The typical approach in these cases, following Rehr and Mermin \cite{Rehr1973}, is to realize the $t,H$ perturbations as suitable linear combinations of the actual variables of the model.\\

In the scaling region, when both the relevant perturbations are present,
all the thermodynamic observables  depend on the scaling combination\footnote{Notice that our definition of $\eta$ differs from that of ref. \cite{fonseca2003ising} by a factor $2\pi$.}
 \begin{equation}
    \eta\equiv \frac{t}{|H|^{\frac{d-x_{\epsilon}}{d-x_{\sigma}}}} = \frac{t}{|H|^{\frac{1}{\beta\delta}}}
    \label{eta1}
\end{equation}

The three limits in which only one of the two perturbations is present ($H=0,t<0$),  ($H\not=0,t=0$) and ($H=0,t>0$) correspond respectively to $\eta=-\infty$, $\eta=0$ and $\eta=+\infty$. In these limits $\Omega$ can be written in terms of the standard universal amplitude ratios
$Q_2$,  ${\Gamma_+}/{\Gamma_-}$ and ${\xi_-}/{\xi_+}$ (see appendix \ref{App:A}) as follows

\begin{equation}
    \begin{aligned}
      \Omega(\eta)&=  1 &\quad & \eta=-\infty  \\
      \Omega(\eta)&= \frac{1}{Q_2} \left(\frac{\Gamma_+}{\Gamma_-}\right) \left(\frac{\xi_-}{\xi_+}\right)^{\gamma / \nu}   &\quad & \eta=0  \\
      \Omega(\eta)&= \left(\frac{\Gamma_+}{\Gamma_-}\right) \left(\frac{\xi_-}{\xi_+}\right)^{\gamma / \nu} &\quad & \eta=+\infty  \\
         \label{eta2}
    \end{aligned}
\end{equation}

These values can be used as benchmarks to test the reliability of our estimates  and as ``anchors" of the reference frame we are constructing.

\section{Parametric Representation}

It is useful to introduce a parametric representation of the critical equation of state, that not only satisfies the scaling hypothesis but additionally allows a simpler implementation of the analytic properties of the equation of state itself. Following \cite{schofield1969}, we express the thermodynamic variables $t, H$ in terms of a couple of parameters $R, \theta$ both positive. Intuitively, the first measures the distance form the critical point in the $(t,H)$ plane while the latter corresponds to the angular displacement along lines of constant $R$ around the critical point. The parametrization of $t$ and $H$ results in a parametric expression of $M$ as well, explicitly

\begin{equation}
    \left\{
    \begin{aligned}
    M &= m_0 R^{\beta} \theta, \\
    t &= R (1-\theta^2), \\
    H &= h_0 R^{\beta \delta} h(\theta)
    \end{aligned}
    \right.
    \label{par1}
\end{equation}

Calling $\theta_0>1$ the smallest positive zero of the function $h(\theta)$, we  see from the system (eq. \ref{par1}) that the domain of interest in the $(R,\theta)$ plane is $0 \le \theta \le \theta_0$ for every $R\ge 0$ and that  $\theta=\theta_0$ corresponds to the $H=0$, $t<0$ axis, $\theta=1$ to the $H\not=0$, $t=0$ axis and $\theta=0$ to the $H=0$, $t>0$ one.

The key point of the whole analysis is the determination of the function $h(\theta)$.
There are some general properties which $h(\theta)$ must satisfy. It
must be analytic in this physical domain in order to satisfy the regularity properties of the critical equation of state, i.e. the so-called Griffiths' analyticity \cite{griffiths1967}. Moreover, it must be an odd function of $\theta$ because of the $\Z_2$ symmetry of the system.
The most general choice is thus a polynomial of the type

\begin{equation}
     h(\theta)=\theta + \sum_{n=1}^{k} h_{2n+1} \theta^{2n+1}
    \label{par3}
\end{equation}
Using standard QFT methods \cite{guida1997,campostrini1999,campostrini2002,Caselle2001}
one can extract the coefficients $h_{2n+1}$ from a high temperature 
expansion of the free energy of the model. Using a variational method it is possible to obtain reliable and stable estimates of the coefficients up to $h_7$ in three dimensions and to $h_{11}$ in two dimensions. 
We can use the known amplitude ratios as benchmarks to evaluate the reliability of these parametric representations.  As we will see $h_7$ will be enough to obtain estimates which in three dimensions agree within the errors with the amplitude ratios. The situation is worse in two dimensions, and this will prompt us to address the 2d case with a different approach. 

A similar parametric representation can be introduced also for the correlation length.  Following \cite{tarko1975} we parametrize the square mass of the underlying QFT i.e. $\xi^{-2}$ as follows

\begin{equation}
    \xi^{-2}= R^{2\nu}a_0(1+c\theta^2)
    \label{par6}
\end{equation}
and similarly, for the second moment correlation length

\begin{equation}
    \xi^{-2}_{2}= R^{2\nu}(a_0)_{2nd}(1+c_2\theta^2)
    \label{par7}
\end{equation}

The constants $c$ and $c_2$ can be fixed using the universal ratios 
$\xi_+/\xi_-$ and $\xi_{2,+}/\xi_{2,-}$ respectively and we use then the $Q_2$ and
 $(Q_2)_{2nd}$ ratios to test the parametric representation. 

Truncating at the quadratic order eq.s \ref{par6} and \ref{par7} could seem a too drastic approximation, but we will see below that in $d=3$ it gives quite good results.  The same is not true in $d=2$ where however, as we anticipated above, we shall use a different approach to evaluate $\Omega(\eta)$.

Using the above results we may construct a parametric representation of $\Omega$
as a function of $\theta$

\begin{equation}
    \Omega(\theta)= \Omega_0 \frac{(1-\theta^2+2\beta\theta^2)(1+c\theta^2)^{\frac{\gamma}{2\nu}}}{2\beta\delta\theta h(\theta)+(1-\theta^2)h'(\theta)}
    \label{par8}
\end{equation}

with 

\begin{equation}
    \Omega_0= \frac{(1-\theta_0^2)h'(\theta_0)}
{(1-\theta_0^2+2\beta\theta_0^2)(1+c\theta_0^2)^{\frac{\gamma}{2\nu}}}
    \label{par9}
\end{equation}\\

and a similar expression for $\Omega_2(\theta)$ with $c\to c_2$.\\

There are a few {\it universal}
features of $\Omega$ that we may deduce from this expression and hold for any possible realization of the Ising universality class

\begin{itemize}
\item
$\Omega$ is a monotonic decreasing function of $\theta$ (see fig. \ref{fig1} for a plot of $\Omega(\theta)$ in $d=3$), therefore it can be inverted. From any given experimental estimate of $\Omega$ a precise value of $\theta$ 
may be extracted.
\item
Using $\Omega$ we may identify the critical isothermal line, which corresponds to $\Omega_{iso}\equiv\Omega(\theta=1)$ {($\Omega_{iso}=1.08376...$ in $d=3$).}
\item
There is a maximum value of $\Omega$ which corresponds to $\Omega_{max}=\Omega(\theta=0)=\Omega_0$ {($\Omega_{max}=1.35502...$ in $d=3$).}  

\end{itemize}

In practical applications one may be interested in the expression of $\Omega$ as a function of $\eta= t/|H|^{\frac{1}{\beta\delta}}$. It is useful to introduce a ``universal" version of the scaling variable defined as
\begin{equation}
    \tilde\eta=\left(h_0\right)^{\frac{1}{\beta\delta}}\eta 
    \label{tildeeta}
\end{equation}

Using the parametric representation of eq. \ref{par1} it is possible to write the expansion of $\theta$ as a function of $\tilde\eta$ in the neighbourhood of the three singular points $\tilde\eta=(\pm\infty,0)$ and then $\Omega$ as a function of $\tilde\eta$, which we plot in fig. \ref{fig2} below in the $d=3$ case (see appendix \ref{App:B} for more details on this expansion).

\subsection{Results in $d=3$}
In three dimensions we have \cite{campostrini1999,campostrini2002}
\begin{align}
    h(\theta)&= \theta+ h_3\, \theta^3+ h_5\, \theta^5 +h_7\, \theta^7 + O(\theta^9)\nonumber\\
    &= \theta-  0.736743 \, \theta^3+ 0.008904 \, \theta^5 - 0.000472\, \theta^7+ O(\theta^9)
    \label{eq4.22}
\end{align}
and the resulting value for the smallest positive root is $\theta_0^2=1.37861$.\\
Inserting these values in eq. \ref{par5} we find  
 $\Gamma_+/ \Gamma_-\sim 4.76$ which is in good agreement (with a difference of the order of 1\%) with the Montecarlo estimate $\Gamma_+/\Gamma_-=4.714(4)$ reported in \cite{Hasenbusch:2010ua,Hasenbusch:2015iat}.

Using the known values of $\xi_+/\xi_-$ and $\xi_{2,+}/\xi_{2,-}$ we then obtain
$c=0.0416$ and  $c_2=0.0776$. To check the reliability of this parametric representation we may use them to predict $Q_2$ and  $(Q_2)_{2nd}$:
we end up with $Q_2\sim 1.250$ and  $(Q_2)_{2nd} \sim 1.209$ which, again, are in agreement (with a difference of the order of 3\%) with the best numerical estimates reported in eq.s \ref{res5}, \ref{res6}.

\begin{figure}[h!]
\centering
\includegraphics[width=0.69\textwidth]{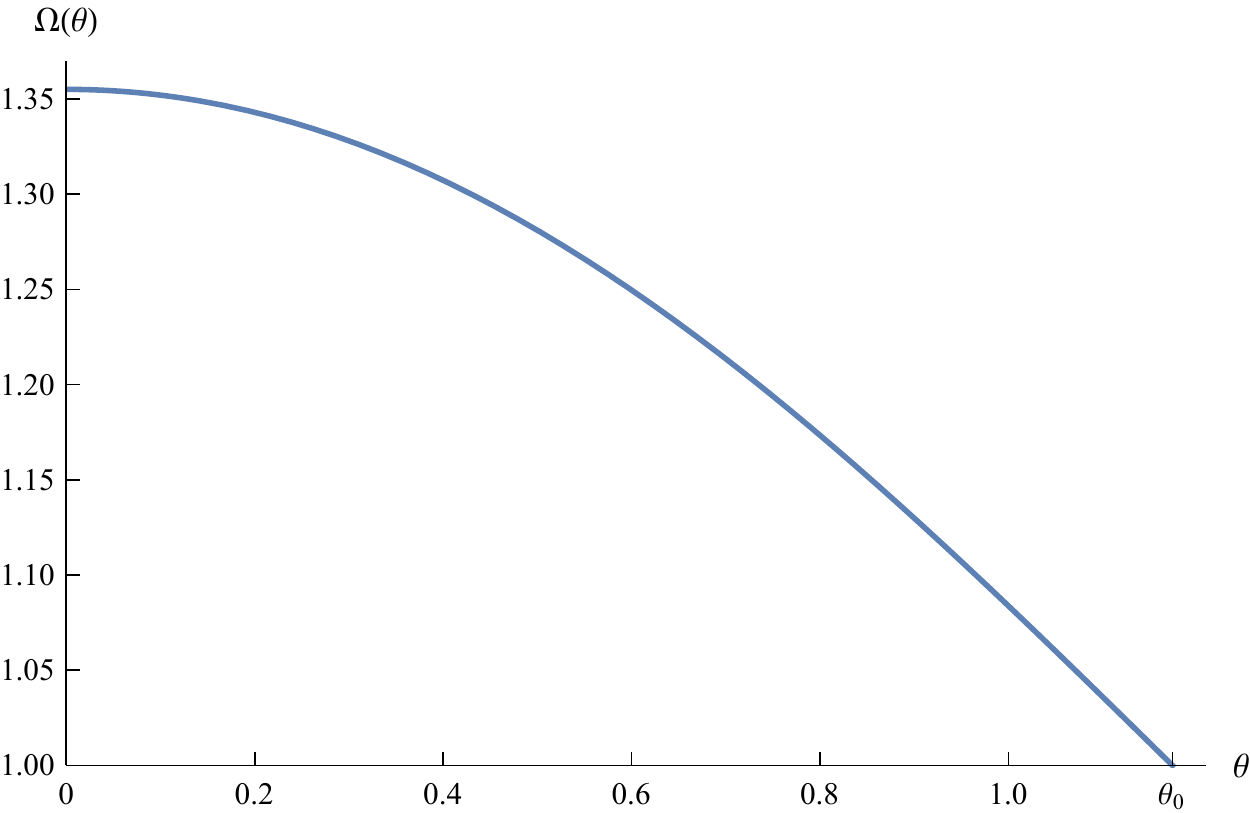}
\caption{Result for $\Omega(\theta)$ according to eq. \ref{par8} with $0\le \theta \le \theta_0$ in $d=3$.}
\label{fig1}
\end{figure}

This tells us that we can trust the parametric representation of $\Omega(\theta)$ discussed above. We plot the result in fig. \ref{fig1}.

\begin{figure}[h!]
\centering
\includegraphics[width=0.69\textwidth]{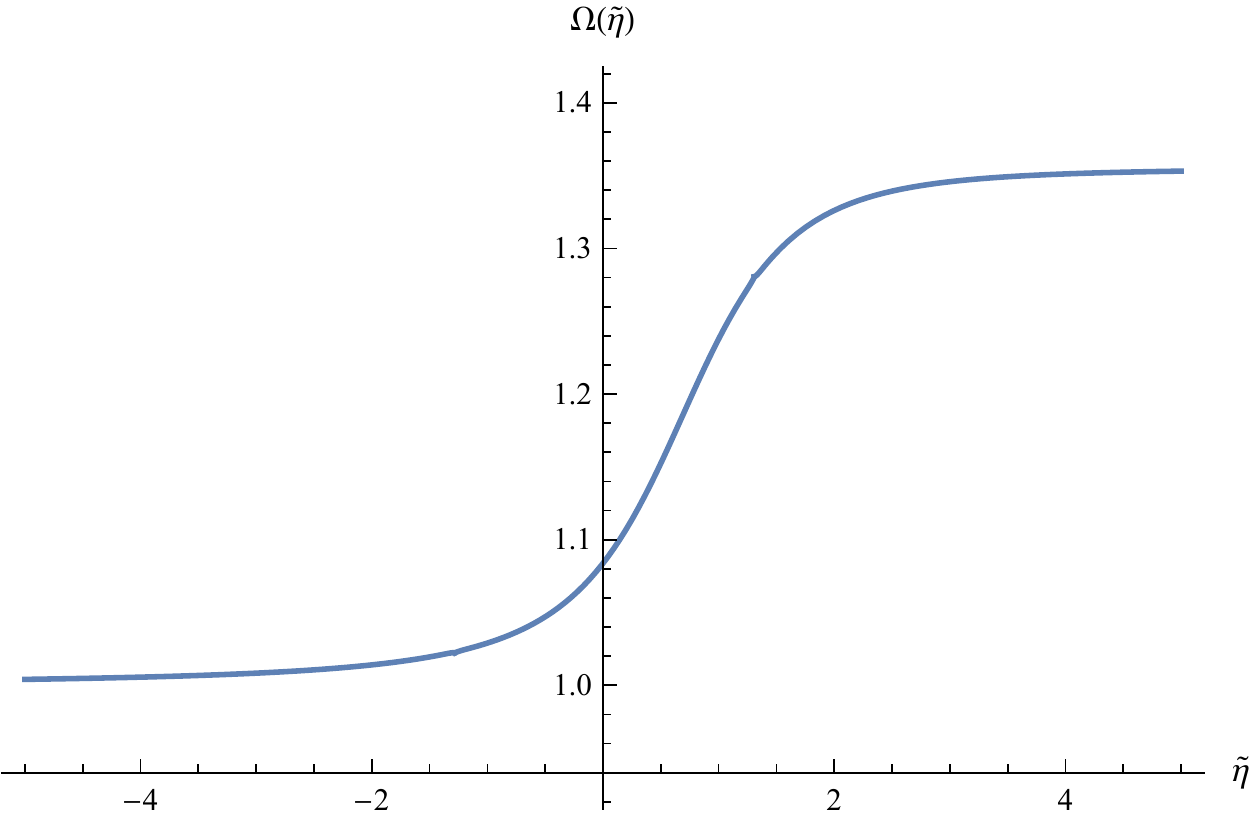}
\caption{Result for $\Omega(\tilde\eta)$ in the three limits of $\tilde\eta$ given in eq.s \ref{eq35}, \ref{eq36}, \ref{eq37} in $d=3$.}
\label{fig2}
\end{figure}

We find in particular {$\Omega_{iso}=1.08376...$ and $\Omega_{max}= 1.35502...$}\\
In fig. \ref{fig2} we plot the expression of $\Omega$ as a function of $\tilde\eta$.

\subsection{Results in $d=2$}
In two dimensions the situation is not as good as in $d=3$.
We have \cite{Caselle2001}
\begin{align}
    h(\theta)&= \theta+ h_3\, \theta^3+ h_5\, \theta^5 +h_7\, \theta^7+ h_9\, \theta^9+ h_{11}\, \theta^{11} + O(\theta^{13})\nonumber\\
    &= \theta- 1.07745 \, \theta^3+ 0.146609 \, \theta^5 + 0.0224263\, \theta^7+ 0.00549457 \, \theta^9 \nonumber\\
    &+ 0.00612906\, \theta^{11} +O(\theta^{13})
    \label{eq4.21}
\end{align}
and the smallest positive zero is $\theta_0^2=1.16951$.\\
From this we find $\Gamma_+/\Gamma_-\sim 39.63$  which is 5\% away from the exact value: $\Gamma_+/\Gamma_-=37.6936520..$  \cite{delfino1998}.
For the correlation length we get, from the exact values of $\xi_+/\xi_-$ and $\xi_{2,+}/\xi_{2,-}$, the following estimates: $c= -0.75678..$ and  $c_2= -0.60933..$.  Plugging these values in the expression for $Q_2$ and  $(Q_2)_{2nd}$ we find for instance $Q_2=5.3342..$ and  $(Q_2)_{2nd}=3.52360$ which are rather far from the exact values $Q_2=3.23513834..$  and $(Q_2)_{2nd}=2.8355305..$ reported in \cite{delfino1998}. This prompted us to address the study of the behaviour of $\Omega(\eta)$ in $d=2$ with a different approach.

\section{Exact expression for $\Omega(\eta)$ in the $d=2$ case}

In $d=2$, one can obtain much more precise results performing a perturbative expansion around the exact solutions along the two axes $H=0$ and $t=0$.
A powerful tool to study the free energy of a perturbed CFT is the well known Truncated Conformal Space Approach (TCSA) \cite{Yurov:1989yu,Yurov:1991my}.  In our case, thanks to the exact mapping of the $H=0$ model to the QFT of a free Majorana fermion
it is possible to construct a more effective version of the TCSA which  uses the free fermions as a basis, the ``Truncated Free-Fermion Space Approach"~\cite{fonseca2003ising}. With this approach it is possible to evaluate the free energy for almost all values of $\eta$~\cite{fonseca2003ising, Fonseca:2003ee,Fonseca:2006au,Zamolodchikov:2013ama} and from that of the susceptibility. With similar methods it is also possible to study the perturbed mass spectrum of the theory ~\cite{Fonseca:2003ee,Fonseca:2006au,Zamolodchikov:2013ama} and hence the correlation length $\xi$ which is 
the inverse of the lowest mass of the spectrum $M_1$. The resulting expansions for $\chi$ and $M_1$, in the three singular limits, are reported in appendix \ref{App:C}.\\

\begin{figure}[h!]
\centering
\includegraphics[width=0.69\textwidth]{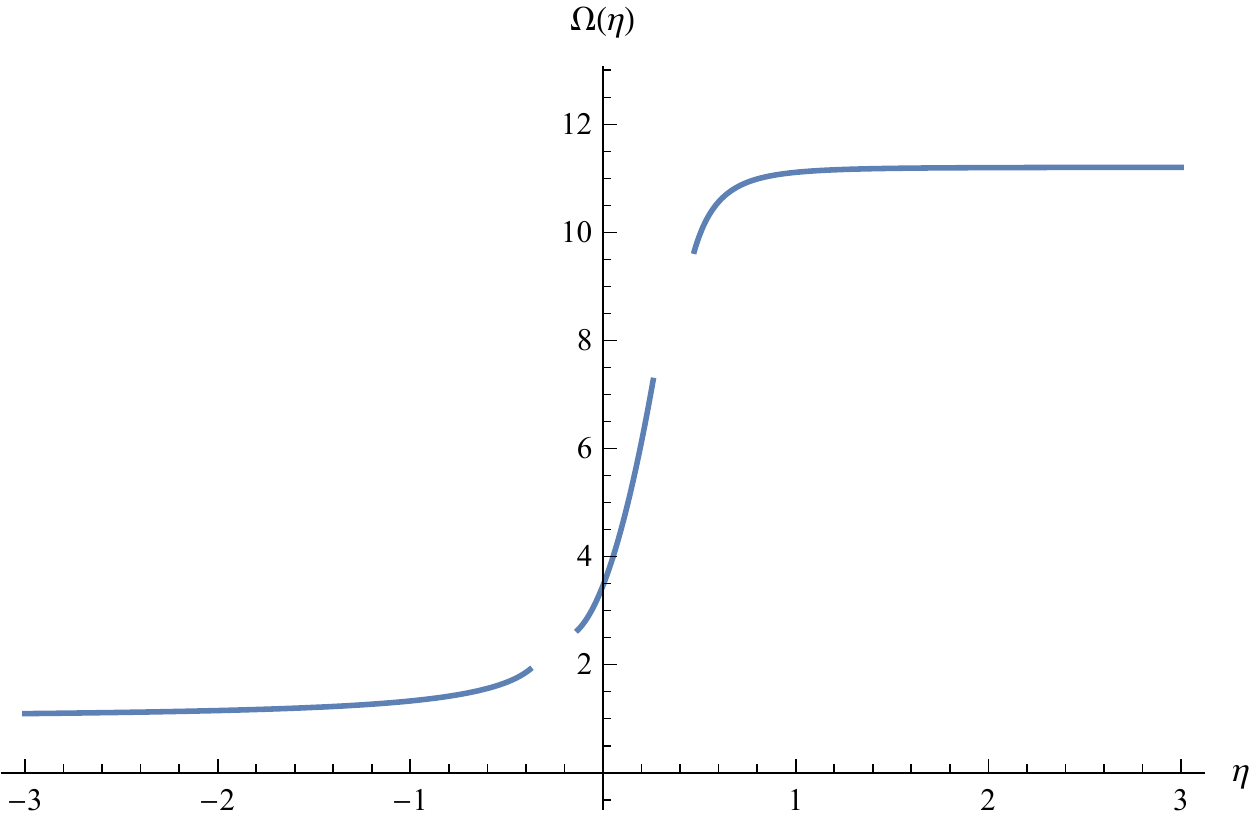}
\caption{Result for $\Omega(\eta)$ in the three limits of $\eta$ given in eq.s \ref{new3} in $d=2$.}
\label{fig3}
\end{figure}

Combining these quantities we obtain a precise estimate for $\Omega$, which we plot in fig. \ref{fig3}. The resulting polynomial expansions in the three regions of interest are

\begin{equation}
    \begin{aligned}
      \Omega(\eta)&= \sum_n \frac{\Omega_n^-}{(-\eta)^{\frac{5}{8}n}} &\quad & \eta \to -\infty  \\
      \Omega(\eta)&= \sum_n \Omega_n^0 ~\eta^n &\quad & \eta \sim 0  \\
      \Omega(\eta)&= \sum_n \frac{\Omega_n^+}{\eta^{\frac{5}{8}n}} &\quad & \eta \to +\infty  
        \label{new3}
    \end{aligned}
\end{equation}

The coefficients are  reported in tab. \ref{tab3}.

\begin{table}[h!]
\centering
\begin{tabular}{l*{3}{|l}}
$n$ & \multicolumn{1}{|c|}{$\Omega_n^-$} & \multicolumn{1}{|c|}{$\Omega_n^0$} & \multicolumn{1}{|c}{$\Omega_n^+$} \\
\hline
0 & $\hphantom{-}1$ & $\hphantom{-}3.46396$ & $\hphantom{-}11.2064$ \\
1 & \multicolumn{1}{|c|}{---} & $\hphantom{-}9.078$ & \multicolumn{1}{|c}{---} \\
2 & $\hphantom{-}0.40031$ & $\hphantom{-}21.23$ & \multicolumn{1}{|c}{---} \\
3 & $-0.075963$ & \multicolumn{1}{|c|}{---} & \multicolumn{1}{|c}{---} \\
4 & $\hphantom{-}0.025182$ & \multicolumn{1}{|c|}{---} & \multicolumn{1}{|c}{---} \\
5 & $-0.0304089$ & \multicolumn{1}{|c|}{---} & \multicolumn{1}{|c}{---} \\
6 & $\hphantom{-}0.006086$ & \multicolumn{1}{|c|}{---} & $-0.09433$ \\
7 & $-0.00191$ & \multicolumn{1}{|c|}{---} & \multicolumn{1}{|c}{---} \\
8 & $\hphantom{-}0.00410$ & \multicolumn{1}{|c|}{---} & \multicolumn{1}{|c}{---} \\
9 & $-0.001035$ & \multicolumn{1}{|c|}{---} & \multicolumn{1}{|c}{---} \\
\end{tabular}
\caption{Expansion coefficients of $\Omega(\eta)$ in the three regimes of interest, according to eq.s \ref{new3}.}
\label{tab3}
\end{table}

As expected, the three limiting cases $\Omega(\pm\infty), \Omega(0)$ (which correspond to the $n=0$ values in the table) 
agree with the universal values obtained plugging in eq. \ref{eta2} the universal ratios quoted in eq. \ref{res2}. 

\section{Three examples of Ising behaviour in finite temperature QCD}

Among the many physical realizations of the Ising universality class in this paper we decided to focus on three examples taken from high energy physics and in particular from the lattice regularization of QCD at finite temperature.

\subsection{The deconfinement transition in the $SU(2)$ pure gauge theory}

The most direct realization of the Ising universality class in Lattice Gauge Theories (LGTs) is given by the deconfinement transition of pure gauge theories with a symmetry group $G$ which has $\Z_2$ as center. This result is based on   
the Svetitsky Yaffe approach~\cite{Svetitsky:1982gs} to the  study of finite temperature $(d+1)$-dimensional pure gauge theories.
The main idea of \cite{Svetitsky:1982gs} is to
construct a $d$-dimensional effective theory from the original one
by integrating out the spacelike links and keeping as only remaining degrees of
freedom the Polyakov loops. These loops are then treated as spins of an effective
$d$-dimensional model whose global symmetry must coincide with the center of the original gauge group. 
If both the phase transitions of the original gauge theory and that of the effective spin model are continuous, they must  belong to the same  universality class and one can use the effective model (which is usually much simpler than the original gauge theory) to extract informations on the deconfinement transition of the original model. 
These results are very general: if in particular we focus on gauge theories with a gauge group $G$ whose center is $\Z_2$ (like for instance, $\Z_2$ itself, $SU(2)$ or $Sp(2N)$), the deconfinement transition will belong to the Ising universality class. A well studied example of this correspondence is the $SU(2)$ model~\cite{Engels:1990vr, Fingberg:1992ju, Engels:1994xj}.  Even if the gauge group is not $SU(3)$ and the model only contains gluonic degrees of freedom this simplified model shares a lot of properties with QCD: the presence of a confining flux tube at low temperatures, a deconfined phase at high temperature, a rich glueball spectrum, asymptotic freedom. For these reasons it has been studied a lot in the past both in (2+1) \cite{KorthalsAltes:1996xp,Liddle:2008kk,Athenodorou:2011rx} and in (3+1) \cite{Athenodorou:2010cs} dimensions. Thanks to the Svetitsky-Yaffe construction, several gauge invariant observables of the $SU(2)$ model can be mapped to equivalent Ising observables
\begin{itemize}
\item
The Polyakov loop is mapped to the spin ($\Z_2$ odd) operator and thus the Polyakov loop susceptibility is mapped to the magnetic susceptibility $\chi$ of the Ising model.
\item
The deconfining transition of the gauge model corresponds to the magnetization transition of the Ising model. In particular,
the confining phase (low temperature of the gauge model) is mapped to the $\Z_2$ symmetric phase (high temperature phase) of the Ising model, while
the deconfined phase is mapped to the broken symmetry phase of the Ising model. 
\item
The Wilson action (i.e. the trace of the ordered product of the gauge field along the links of a plaquette) is mapped to the energy operator\footnote{Actually it is mapped to the most general $\Z_2$ even Ising observable, which is a mixture of the identity and energy operators, but the identity operator plays no role in this context.}  of the Ising model. 
\item
The screening mass of the gauge model in the deconfined phase is mapped to the mass of the Ising model in the low temperature phase, while 
$N_t \sigma$, where $\sigma$ is the string tension and $N_t$ is the inverse temperature of the gauge model (i.e. the size $N_t$ of the lattice in the compactified direction which defines the finite temperature setting in LGTs)  is mapped to the inverse of the high temperature correlation length of the Ising model.
\end{itemize}
This mapping has been widely used in the past years to
predict the behaviour of various physical observables of the gauge model near the deconfinement transition, like for instance the short distance behaviour of the Polyakov loop correlator~\cite{Caselle:2019tiv}, the width of the flux tube~\cite{Caselle:2012rp}, the Hagedorn-like behaviour of the glueball spectrum~\cite{Caselle:2015tza} or the behaviour of the universal $\xi/\xi_{2}$ ratio~\cite{Caselle:2017xrw}. More generally the mapping
allows one to relate all the thermodynamic observables of the two models and in particular also $\Omega(\eta)$, which can thus be evaluated in the $SU(2)$ LGT and then compared with the QFT prediction discussed in the previous section.

\subsection{QCD with dynamical quarks and the Columbia plot}
\begin{figure}[h!]
\centering
\includegraphics[width=0.59\textwidth]{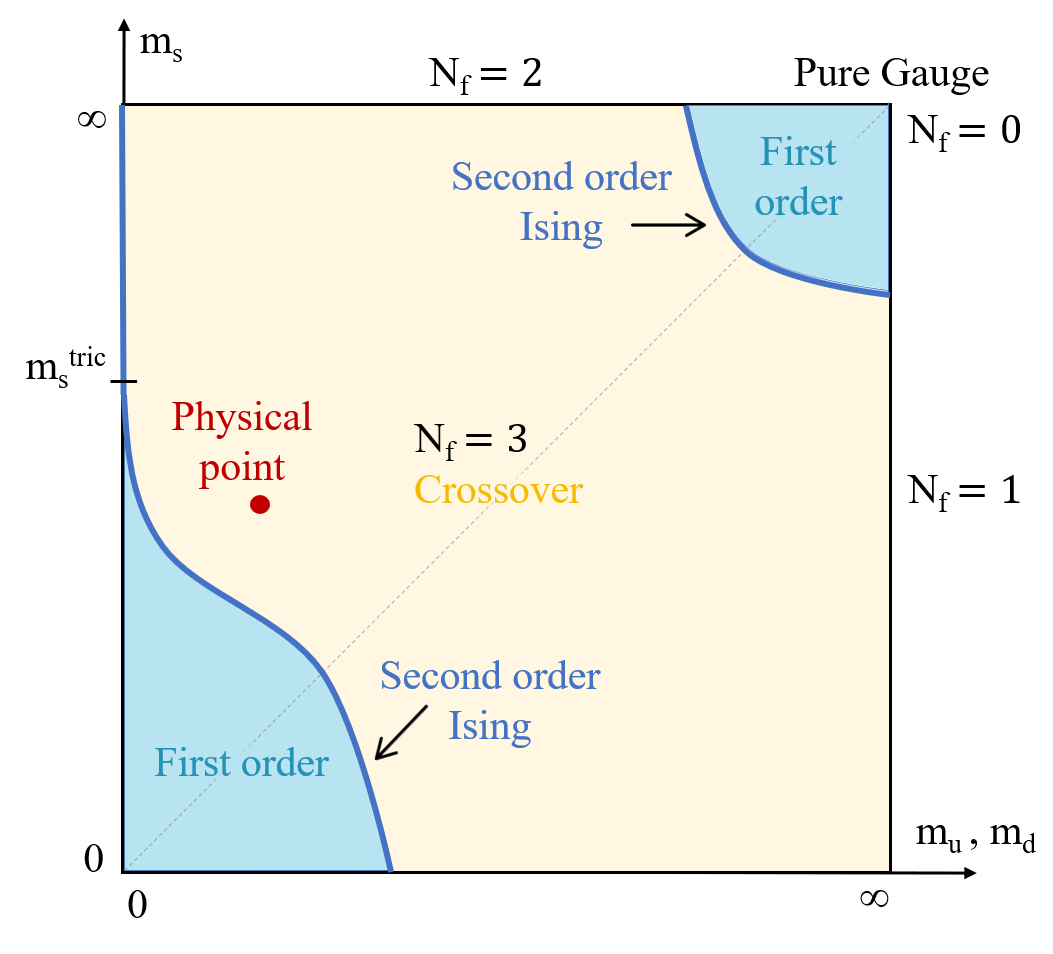}
\caption{Columbia Plot.}
\label{fig:CP}
\end{figure}

The situation is different if we study full QCD, i.e. if we include dynamical quarks in the model. In this case  the center symmetry is explicitly broken by the Dirac operator and all the above considerations do not hold anymore. For physical values of the quark masses there is no phase transition between the high temperature quark-gluon plasma phase and the low temperature confined phase which are only separated by a  smooth crossover~\cite{Aoki:2006br, Bazavov:2011nk}.
However in the phase diagram of the model one finds a rich structure of phase transitions as the values of the masses of the quarks 
are varied \cite{Pisarski:1983ms}.  This pattern of phase transitions is summarized in the well known Columbia plot, which we report here in fig. \ref{fig:CP}. The plot describes the nature of the phase transitions as a function of the quark masses. On the two axes are respectively the masses of up-down quarks (x axis) and the mass of the strange quark (y axis). We see in the central part of the plot a wide region, where the physical point lies, in which there is no phase transition but only a crossover between the two phases. In the top right and in the bottom left corners there are two regions where the transition is of the first order.  These regions end with two lines of second order transitions which are expected to both be of the Ising type. 

The phase diagram reported in the Columbia Plot can be studied with standard Montecarlo simulations and in the vicinity of the critical lines the results of these simulations could be mapped using our tools to the Ising phase diagram and then compared with the Ising predictions as we discussed above for the simpler case of the pure gauge $SU(2)$ model.

It is important to notice that the two critical regions are of different nature.  The one in the top-right corner is a deconfinement transition similar to the one discussed in the previous section.  In fact in the limit of infinite mass quarks the model becomes a pure gauge theory.  In this limit the $SU(3)$  gauge model has a first order deconfinement transition (differently from the $SU(2)$ one discussed above which is of second order). As the mass of the quarks decreases the gap in the order parameter decreases and the first order region ends into a critical line of Ising-like phase transitions. 

The critical region in the bottom left portion of the Columbia Plot has a completely different origin. It describes the restoration of the
chiral symmetry in QCD at finite temperature  and small quark masses.
In QCD with three massless quark flavours the chiral phase transition is expected to be first order and to remain of first order even for small but non-zero values of the quark masses. As the quark masses increase, the gap in the order parameter decreases and the first order region terminates in a critical line of the Ising type. Above this line chiral symmetry is restored through a smooth crossover.
In this case, the $\Z_2$ symmetry is an ``emergent" symmetry and the identification of  the $H$ and $t$ axes of the equivalent 3d Ising model is not trivial (see a discussion on this issue in 
\cite{Karsch:2001nf,Bazavov:2017xul} and in sect. \ref{QCD} below) and thus a universal charting of the scaling region could indeed be useful.

Even if the precise location of the critical line is still debated, it seems that the 
physical point is not too far from this bottom left Ising line. If this is the case, then our analysis could be applied, 
maybe with some degree of approximation, also to the physical point.

\subsection{The critical ending point of the QCD phase diagram at finite chemical potential}
\label{QCD}
The most interesting application of our results is for QCD  at finite baryon density, which is realized by adding a finite chemical potential $\mu$  to the QCD lagrangian.  This regime is particularly interesting since it can be explored experimentally in heavy-ion collisions 
\cite{Lacey:2006bc,Stephanov:1998dy,Adamczyk:2013dal, Lacey:2014wqa, Gazdzicki:2015ska, Adare:2015aqk, Adamczyk:2017wsl, Yin:2018ejt} and at the same time it cannot be studied using Montecarlo simulations due to the well known sign problem (see for instance~\cite{deForcrand:2010ys, Gattringer:2016kco} for a discussion of the sign problem in this context).

\begin{figure}[h!]
\centering
\includegraphics[width=0.749\textwidth]{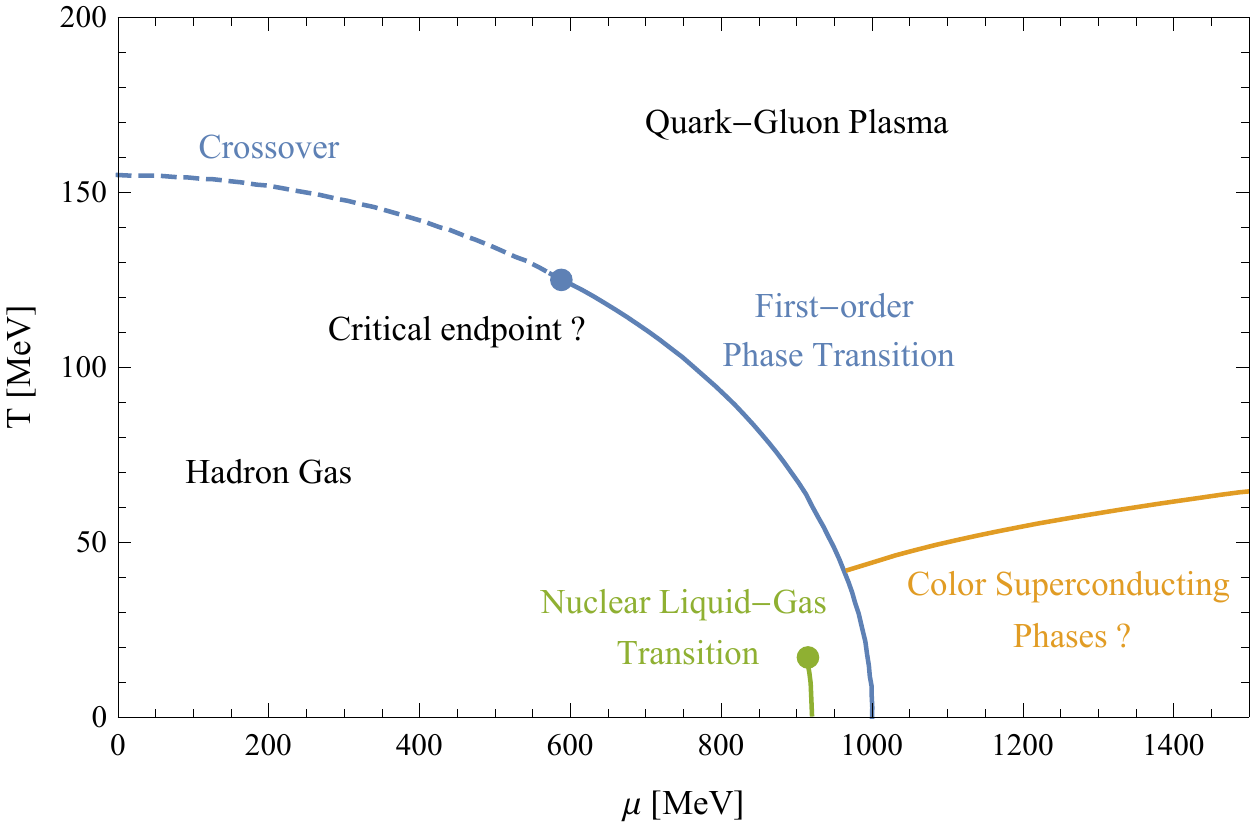}
\caption{QCD phase diagram at finite chemical potential.}
\label{fig:QCD}
\end{figure}

In this regime the QCD phase diagram is expected to reveal interesting novel phases~\cite{Alford:1998mk, Stephanov:2007fk}. In particular it is widely expected that the hadronic phase  (low $T$, low $\mu$) should be separated from the quark-gluon plasma phase 
(high $T$, high $\mu$) by a line of first order transitions with a critical endpoint at finite critical values of $T$ and $\mu$ (see fig. \ref{fig:QCD})
 which should again belong to the Ising universality class ~\cite{Halasz:1998qr, Berges:1998rc, Hatta:2002sj,Nonaka:2004pg, Antoniou:2018ico, Parotto:2018pwx, Pradeep:2019ccv, Martinez:2019bsn}.   

Also for this model, as for the liquid-vapour transition (or the chiral transition discussed above), the $\Z_2$ symmetry is not realized explicitly but
it is just an ``emergent'' symmetry. This class of models is typically addressed with the ``mixing-of-coordinates" scheme proposed in \cite{Rehr1973}. The approach was pursued for the finite density QCD case in \cite{Nonaka:2004pg} and \cite{Parotto:2018pwx}, leading to very interesting results. In both cases the mapping between Ising and QCD coordinates was performed via the parametric representation of the equation of state. In this respect, the explicit expression of $\Omega$ in terms of $\theta$ that we discussed in this paper could be used as a shortcut in the process and could facilitate the identification of Ising-like behaviours in the experimental data.

As more and more experimental results are obtained, it will become possible to directly test them with universal predictions from the Ising model and it will be important to have a precise charting of the Ising phase diagram to organize results and drive our understanding of strongly coupled QCD in this regime. Our paper is a first step in this direction.  We proposed and studied one particular combination, chosen for its simplicity from a theoretical point of view and its accessibility from an experimental and  numerical point of view, but other combinations are  possible and could be studied using, as we suggest here, parametric representations in $d=3$ and/or expansion around the exactly integrable solutions in $d=2$.

\section*{Acknowledgments}
We thank C. Bonati, M. Hasenbusch and M. Panero for 
useful comments and suggestions.

\begin{appendix}

\section{Critical amplitudes and universal amplitude ratios}
\label{App:A}
 We list below the scaling behaviour of  the observables used in the main text
\begin{equation}
    \begin{aligned}
      \xi &\approx \xi_+\, t^{-\nu} &\quad \xi_{2} &\approx \xi_{2,+}\, t^{-\nu}&\quad t>0, H=0 \\
 \xi &\approx \xi_-\, (-t)^{-\nu}&\quad \xi_{2} &\approx \xi_{2,-}\, (-t)^{-\nu} &\quad t<0, H=0 \\
 \xi &\approx \xi_c\, |H|^{-\nu_c}&\quad \xi_{2}  &\approx \xi_{2,c}\, |H|^{-\nu_c} &\quad t=0, H\neq 0
        \label{def6}
\nonumber
    \end{aligned}
\end{equation}

\begin{equation}
    \begin{aligned}
        \chi &\approx \Gamma_+\, t^{-\gamma} &\quad   & &\quad t>0, H=0   \\
        \chi &\approx \Gamma_-\, (-t)^{-\gamma} &\quad M &\approx B\, (-t)^{\beta}&\quad t<0, H=0 \\
        \chi &\approx \Gamma_c\, |H|^{-\gamma_c} &\quad M &\approx B_c\, |H|^{\frac1\delta} &\quad t=0, H\neq 0
        \label{def6bis}
\nonumber
    \end{aligned}
\end{equation}

where the critical indices are defined in terms of the scaling exponents as follows
\begin{equation}
    \begin{gathered}
\beta =\frac{x_{\sigma}}{(d-x_{\epsilon})} \qquad \delta= \frac{(d-x_{\sigma})}{x_{\sigma}} \\  \gamma= \frac{(d-2 x_\sigma)}{d-x_\epsilon} \qquad \nu =\frac{1}{d-x_\epsilon} \\ \gamma_c=\frac{(d-2 x_{\sigma})}{(d-x_{\sigma})} \qquad \nu_c=\frac{1}{(d-x_{\sigma})}
    \end{gathered}
    \label{def7}
\end{equation}

From these definitions it is possible to construct the following universal amplitude ratios
\begin{equation}
\frac{\Gamma_+}{\Gamma_-},\quad\frac{\xi_+}{\xi_-},\quad
\quad  Q_2=\left(\frac{\Gamma_+}{\Gamma_c} \right)\left(\frac{\xi_c}{\xi_+} \right)^{\gamma/\nu}  \label{def8}
\end{equation}

\begin{equation}
\quad\frac{\xi_{2,+}}{\xi_{2,-}},\quad \quad  (Q_2)_{2nd}=\left(\frac{\Gamma_+}{\Gamma_c} \right)\left(\frac{\xi_{2,c}}{\xi_{2,+}} \right)^{\gamma/\nu}  \label{def9}
\end{equation}

\subsection{Exact results for the amplitude ratios in $d=2$} 
In $d=2$, 
thanks to the exact integrability of the two relevant perturbations all the above universal quantities are known exactly \cite{delfino1998}

\begin{equation}
    x_{\epsilon}=1 \qquad x_{\sigma}=\frac{1}{8}
    \label{res1}
\end{equation}

\begin{equation}
\frac{\Gamma_+}{\Gamma_-}=37.6936520..,\quad\frac{\xi_+}{\xi_-}= 2, \quad  Q_2=3.23513834...  \label{res2}
\end{equation}

\begin{equation}
\quad\frac{\xi_{2,+}}{\xi_{2,-}}= 3.16249504..,\quad \quad  (Q_2)_{2nd}=2.8355305... 
\label{res3}
\end{equation}

\subsection{Numerical estimates of the amplitude ratios in $d=3$}

In three dimensions there are no exact results but, thanks to the recent improvement of the  bootstrap approach~\cite{Kos:2016ysd,berg1979,Komargodski:2016auf} and to the remarkable precision of recent Montecarlo simulations \cite{Hasenbusch:2010ua,Hasenbusch:2015iat,engels2003}, reliable numerical estimates for all these quantities exist

\begin{equation}
    x_{\epsilon}=1.412625(10) \qquad x_{\sigma}=0.5181489(10)
    \label{res4}
\end{equation}

\begin{equation}
\frac{\Gamma_+}{\Gamma_-}=4.714(4),\quad\frac{\xi_+}{\xi_-}= 1.896(3), \quad  Q_2=1.207(2)  \label{res5}
\end{equation}

\begin{equation}
\quad\frac{\xi_{2,+}}{\xi_{2,-}}= 1.940(2),\quad \quad  (Q_2)_{2nd}=1.179(2)  
\label{res6}
\end{equation}

It is also possible to choose realizations of the model in which the amplitude of the first irrelevant operator is tuned toward zero, thus allowing a more efficient approach to the fixed point.  This is for instance the idea followed in \cite{Hasenbusch:1998gh,Hasenbusch:1999mw,Hasenbusch:2010ua,Hasenbusch:2011yya,Hasenbusch:2015iat,Hasenbusch:2017yxr,engels2003} to improve the numerical estimates of universal quantities using Montecarlo simulations.

\section{Useful results in the parametric representation}
\label{App:B}

\subsection{Fixing the non-universal constant $h_0$}

The scaling parameter $\eta$ is defined modulo a non-universal constant $h_0$ which sets its scale and depends on the specific model at hand, i.e. on the specific realization of the Ising universality class in which one is interested.
However, given such a realization, it is rather easy to fix $h_0$. The simplest option is to measure (numerically or experimentally) the magnetization $M$ and the susceptibility $\chi$ along two directions (or in the same direction) in the $(t,H)$ plane, for instance along the critical line of first order phase transitions. Then from the ratio of the two amplitudes one can easily extract $h_0$. We report here for completeness the corresponding expressions in the case in which one measures, besides the amplitude $B$ of the magnetization,  the value of $\Gamma_+$ or that of $\Gamma_-$

\begin{equation}    
h_0=\frac{B}{\Gamma_+}\frac{(\theta_0^2-1)^\beta}{\theta_0}
    \label{h0}
\end{equation}

or
\begin{equation}    
h_0=-\frac{B}{\Gamma_-}\frac{(\theta_0^2-1)^{\gamma+\beta-1}(1-\theta_0^2+2\beta\theta_0^2)}{\theta_0 h'(\theta_0)}
    \label{h0bis}
\end{equation}

For instance, in the case of the 3d Ising model defined by eq.s \ref{def1}, \ref{def2} we have \cite{Caselle:1997hs} $B= 1.6920(5)$ and $\Gamma_-=0.2394(5)$ to which corresponds $h_0\sim 0.923$
while in the case of the model tuned so as to eliminate the first irrelevant operator discussed above we find 
\cite{Hasenbusch:2015iat} $B=1.9875(3)$ and $\Gamma_+=0.14300(5)$ to which corresponds $h_0\sim 8.62$.

\subsection{Parametric representation of the magnetic susceptibility}

From the parametric representation of the critical equation of state (eq. \ref{par1}) we may obtain the magnetic susceptibility as follows

\begin{equation}
    \chi(R,\theta)=\left(\frac{m_0}{h_0}\right) R^{-\gamma} \frac{1-\theta^2+2\beta \theta^2}{2\beta \delta \theta h(\theta) + (1-\theta^2) h'(\theta)}
    \label{par4}
\end{equation}

and from this expression it is easy to extract the amplitude ratio

\begin{equation}
    \frac{\Gamma_+}{\Gamma_-}= -\frac{ (\theta_0^2-1)^{1-\gamma} h'(\theta_0)}{(1-\theta_0^2+2\beta \theta_0^2)}
    \label{par5}
\end{equation}

\subsection{Expansion of $\theta$ as a function of $\tilde\eta$ in the neighbourhood of the three singular points $\tilde\eta=(\pm\infty,0)$}

We report here the first few terms:
in the  $\theta\to \theta_0$ limit
\begin{align}
    \theta(\tilde\eta) &=\theta_0 + \frac{(\theta_0^2-1)^{\beta\delta}}{h'(\theta_0)} \left(\frac{1}{-\tilde\eta}\right)^{\beta\delta}  + \biggl[ \frac{2\beta\delta\theta_0 (\theta_0^2-1)^{2\beta\delta}}{(\theta_0^2-1) (h'(\theta_0))^2}+\nonumber\\
&- \frac{(\theta_0^2-1)^{2\beta\delta}h''(\theta_0)}{2 (h'(\theta_0))^3} \left(\frac{1}{-\tilde\eta}\right)^{2\beta\delta} +O\left[\left(\frac{1}{-\tilde\eta}\right)^{3\beta\delta}\right],
    \label{eq5.39}
\end{align}

in the  $\theta\to 1$ limit
\begin{align}
    \theta(\tilde\eta)&=1- \frac{1}{2}\left(h(1)\right)^{\frac{1}{\beta\delta}} \tilde\eta + \frac{2 h'(1)-\beta\delta h(1)}{8\beta\delta h(1)} \left(h(1)\right)^{\frac{2}{\beta\delta}} \tilde\eta^2 + O\left(\tilde\eta^3\right),
    \label{eq5.40}
\end{align}

and finally in the $\theta\to 0$ limit
\begin{equation}
    \theta(\tilde\eta)=0+ \left(\frac{1}{\tilde\eta}\right)^{\beta\delta} - \frac{h''(0)}{2} \left(\frac{1}{\tilde\eta}\right)^{2\beta\delta} +O\left[\left(\frac{1}{\tilde\eta}\right)^{3\beta\delta}\right]
    \label{eq5.38}
\end{equation}

\subsection{Expansions of $\Omega$ as a function of  $\tilde\eta$}

Plugging the above expansions into the expression for $\Omega(\theta)$ we find\footnote{{We report here only the first term for the three expansions to avoid too complex formulas, it is straightforward to obtain the next orders.}} for $\Omega(\tilde\eta)$:
in the  $\tilde\eta\to -\infty$ limit
\begin{align}
    \Omega(\tilde\eta) &=1 + \frac{\theta_0 (\theta_0^2-1)^{\beta\delta}}{h'(\theta_0)} \biggl[ \frac{c \gamma}{\nu (1+c\theta_0^2)} - \frac{4\beta}{\left[1+\theta_0^2(2\beta-1)\right] (\theta_0^2-1)}+ \nonumber\\
&+ \frac{2\beta\delta}{(\theta_0^2-1)} - \frac{h''(\theta_0)}{\theta_0 h'(\theta_0)} \biggl] \left(\frac{1}{-\tilde\eta}\right)^{\beta\delta}  + O\left[\left(\frac{1}{-\tilde\eta}\right)^{2\beta\delta}\right],
    \label{eq36}
\end{align}

in the  $\tilde\eta\to 0$ limit
\begin{align}
    \Omega(\tilde\eta)&=\frac{\Omega_0 (1+c)^{\frac{\gamma}{2\nu}}}{\delta h(1)} + \frac{\Omega_0 (1+c)^{\frac{\gamma}{2\nu}}}{2\delta h(1)} \biggl[\frac{h'(1) \left(\beta\delta-1\right)-\delta h(1)\left(\beta-1\right)}{\beta\delta h(1)} +\nonumber\\
&- \frac{c \gamma}{\nu (1+c)} \biggl]\left(h(1)\right)^{\frac{1}{\beta\delta}} \tilde\eta + O(\tilde\eta^2),
    \label{eq37}
\end{align}

and finally in the $\tilde\eta\to +\infty$ limit
\begin{equation}
    \Omega(\tilde\eta)=\Omega_0 - \Omega_0 h''(0) \left(\frac{1}{\tilde\eta}\right)^{\beta\delta} + O\left[\left(\frac{1}{\tilde\eta}\right)^{2\beta\delta}\right]
    \label{eq35}
\end{equation} 

Upon substitution $c\to c_2$ the same results are valid for $\Omega_2(\tilde\eta)$.

\section{Exact results in $d=2$}
\label{App:C}

We report here the expansions, in the three regimes of interest, for $\chi$ and for $M_1\equiv 1/\xi$, obtained using the results reported in \cite{fonseca2003ising, Fonseca:2003ee,Fonseca:2006au,Zamolodchikov:2013ama}.

\begin{equation}
    \begin{aligned}
      \chi(\eta)&= (-t)^{-\frac74} \sum_n \frac{\chi_n^-}{(-\eta)^{\frac{15}{8}n}} &\quad & \eta \to -\infty  \\
      \chi(\eta)&= |H|^{-\frac{14}{15}}\sum_n \chi_n^0 ~\eta^n &\quad & \eta \sim 0  \\
      \chi(\eta)&= t^{-\frac74} \sum_n \frac{\chi_n^+}{\eta^{\frac{15}{8} n}} &\quad & \eta \to +\infty
        \label{new1}
    \end{aligned}
\end{equation}

\begin{table}[h]
\centering
\begin{tabular}{l*{3}{|l}}
$n$ & \multicolumn{1}{|c|}{$\chi_n^-$} & \multicolumn{1}{|c|}{$\chi_n^0$} & \multicolumn{1}{|c}{$\chi_n^+$} \\
\hline
0 & $\hphantom{-}0.00392687$ & $\hphantom{-}0.0851721$ & $\hphantom{-}0.14800123$ \\
1 & $-2.98296\times10^{-4}$ & $\hphantom{-}0.498574$ & \multicolumn{1}{|c}{---} \\
2 & $\hphantom{-}3.34904\times10^{-5}$ & $\hphantom{-}1.67552$ & $-0.00407428$ \\
3 & $-4.7920\times10^{-6}$ & $\hphantom{-}3.33185$ & \multicolumn{1}{|c}{---} \\
4 & $\hphantom{-}8.2208\times10^{-7}$ & $\hphantom{-}0.90703$ & $\hphantom{-}1.1818\times10^{-4}$ \\
5 & $-1.635\times10^{-7}$ & $-20.939$ & \multicolumn{1}{|c}{---} \\
6 & $\hphantom{-}3.694\times10^{-8}$ & $-85.959$ & $-3.4330\times10^{-6}$ \\
7 & $-9.335\times10^{-9}$ & $-169.4$ & \multicolumn{1}{|c}{---} \\
8 & $\hphantom{-}2.610\times10^{-9}$ & $-10.32$ & $\hphantom{-}9.951\times10^{-8}$ \\
9 & $-7.995\times10^{-10}$ & $\hphantom{-}1162$ & \multicolumn{1}{|c}{---} \\
10 & $\hphantom{-}2.67\times10^{-10}$ & $\hphantom{-}4288$ & $-2.87\times10^{-9}$ \\
11 & $-9.62\times10^{-11}$ & \multicolumn{1}{|c|}{---} & \multicolumn{1}{|c}{---} \\
12 & $\hphantom{-}3.74\times10^{-11}$ & \multicolumn{1}{|c|}{---} & $\hphantom{-}8.34\times10^{-11}$ \\
\end{tabular}
\caption{Expansion coefficients of the magnetic susceptibility in the three regimes of interest, according to eq.s \ref{new1}.}
\label{tab1}
\end{table}

\begin{equation}
    \begin{aligned}
      \frac{1}{\xi(\eta)} &= (-t) \sum_n \frac{m_n^-}{(-\eta)^{\frac{5}{4}n}} &\quad & \eta \to -\infty  \\
           \frac{1}{\xi(\eta)} &= |H|^{\frac{8}{15}}\sum_n m_n^0 ~\eta^n &\quad & \eta \sim 0  \\
           \frac{1}{\xi(\eta)} &= t \sum_n \frac{m_n^+}{\eta^{\frac{5}{4}n}} &\quad & \eta \to +\infty  
        \label{new2}
    \end{aligned}
\end{equation}

where we have already substituted the known values of the critical indices of the model. The coefficients are reported in tab.s \ref{tab1}, \ref{tab2}.

\begin{table}[h]
\centering
\begin{tabular}{l*{3}{|l}}
$n$ &  \multicolumn{1}{|c|}{$m_n^-$} & \multicolumn{1}{|c|}{$m_n^0$} & \multicolumn{1}{|c}{$m_n^+$} \\
\hline
0 & $\hphantom{-}4\pi$ & $\hphantom{-}4.404909$ & $\hphantom{-}2\pi$ \\
1 & $\hphantom{-}2.8746$ & $-8.137008$ & \multicolumn{1}{|c}{---} \\
2 & $-0.06576$ & $\hphantom{-}7.908$ & \multicolumn{1}{|c}{---} \\
3 & $-0.00156$ &  \multicolumn{1}{|c|}{---} & $\hphantom{-}0.0686$ \\
4 & $\hphantom{-}0.00438$ & \multicolumn{1}{|c|}{---} & \multicolumn{1}{|c}{---} \\
\end{tabular}
\caption{Expansion coefficients of the correlation length in the three regimes of interest, according to eq.s \ref{new2}.}
\label{tab2}
\end{table}

\end{appendix}

\providecommand{\href}[2]{#2}\begingroup\raggedright\endgroup

\end{document}